\documentclass[twocolumn,superscriptaddress,secnumarabic,amssymb,amsfont,amsmath,nobibnotes,  longbibliography]{revtex4-2}
\usepackage{bm}
\usepackage{mathtools}
\usepackage{graphicx}
\usepackage{float}

\begin{document}
	
	\title{Sensing Vibrations using Quantum Geometry of Electrons}%
	\author{Bhuvaneswari R}%
	\email{bhuvana@jncasr.ac.in}
	\affiliation{Theoretical Sciences Unit, Jawaharlal Nehru Centre for Advanced Scientific Research, Bangalore, 560064, India.}
	\affiliation{School of Electrical \& Electronics Engineering, SASTRA Deemed University, Thanjavur, 613401, India.}
	\author{Mandar M Deshmukh}%
	\email{mandar.m.deshmukh@gmail.com}
	\affiliation{Department of Condensed Matter Physics and Materials Science, Tata Institute of Fundamental Research, Mumbai, 400005, India.}
	\author{Umesh V Waghmare}%
	\email{waghmare@jncasr.ac.in}
	\affiliation{Theoretical Sciences Unit, Jawaharlal Nehru Centre for Advanced Scientific Research, Bangalore, 560064, India.}
	\date{\today}%
	
	\begin{abstract}
		Magnetic field emerging from the geometric curvature of quantum structure of electrons in a crystal bends electronic trajectory causing \textit{anomalous} linear and nonlinear electrical Hall effects that have been observed in \textit{low} symmetry crystals with narrow electronic band gap. We present first-principles theoretical analysis to show that dynamical lowering of crystal symmetry by lattice vibrations results in oscillations in the quantum geometry of electrons which have observable nonlinear Hall signatures. Using these, we introduce a vibrational spectroscopy based on Geometry of Quantum Electronic Structure (GQuES) making specific predictions for the transport and radiative GQuES spectra of 2D materials. As the crystal symmetry is commonly lowered by dynamical fields, GQuES is applicable to a wide range of materials and excitations spanning sub-GHz, THz and infrared frequencies.
	\end{abstract}%
	\maketitle
	
	When the external parameters or fields interacting with a quantum system change slowly along a cyclic path or a loop, its quantum energy state time-evolves adiabatically picking up two phase factors: (i) the dynamical phase factor determined by the energies of quantum states traversed during the cyclic evolution and (ii) an additional factor that depends only on the geometry of quantum states accessed along the loop in the parameter space, as shown by Berry in 1984 \cite{berry1984quantal}. When these parameters correspond to dynamical variables of slow degrees of freedom, the geometric or Berry phases are physically relevant to measurable properties significant to applications \cite{xiao2010berry}. Manifestation of Berry phases is omnipresent and has been realized in diverse phenomena in quantum chemistry, physics and material science \cite{xiao2010berry}. In the last two decades, Berry phases and curvature have been shown to govern electronic topology of crystals that defines quantum states of matter like topological insulators, Dirac and Weyl semimetals \cite{hasan2010colloquium,armitage2018weyl}. 
	
	Bending of electronic trajectory in a crystal due to geometric (Berry) curvature of its quantum electronic structure causes electrical Hall effect even in the absence of magnetic field \cite{karplus1954hall,xiao2010berry}. Such linear anomalous Hall effect \cite{ye1999berry}, observable only in the systems with broken time reversal symmetry, rarely manifests in nonmagnetic systems. Electronic Berry curvature dipole, a first moment of Berry curvature, was shown theoretically to result in the nonlinear Hall (NLH) effect \cite{sodemann2015quantum} which have been observed in the time reversal symmetric systems \cite{ma2019observation}. This highlighted its ability to probe quantum geometry of materials, particularly those with narrow bandgap and low crystallographic symmetry \cite{ma2021topology}. 
	
	In this work, we demonstrate that dynamical excitations lower the symmetry of a crystal and modulate the Berry curvature dipole and consequent signals in its frequency dependent Hall response constitute a powerful “Geometry of Quantum Electronic Structure (GQuES)” based spectroscopic tool to measure their excitation energies. Using first-principles theory and simulations, we illustrate the GQuES tool with specific predictions for experimental measurements of Hall transport and THz/IR emission to measure frequencies of acoustic and optic phonons. Originating from the oscillations in Berry curvature dipole induced by symmetry lowering vibrations, GQuES signals are observable even in centrosymmetric nonmagnetic systems with vanishing intrinsic Berry curvature. Applicability of GQuES vibrational spectroscopy is shown to extend to a wide band gap, inert crystal like 2D h-BN, when aligned with graphene as a host of nontrivial quantum geometry. It measures frequencies of acoustic and optic phonons, spanning from sub-GHz, THz to IR frequencies.

	In the theoretical demonstration here, we choose 2D crystals. An ultra-thin, two-dimensional (2D) crystal retains its intrinsic carrier mobility when used as a channel in a field effect transistor, as a consequence of the dangling-bond free individual layers, effectively reducing the surface scattering \cite{liu2021promises}. A 2D crystal with strong intralayer covalent bonding and weak interlayer van der Waals (vdW) forces, offers superior electrostatic control over its charge carriers through the gate voltage \cite{wang2022road}. 2D crystals with distinct electronic properties can be integrated \cite{liu2019van} to enable multi-functionality and ease in fabrication of low powered electronic devices with enhanced performance \cite{liu2020two}. Noting the advantage in experimental realization and interesting quantum geometry exhibited by some of the 2D materials \cite{ma2019observation}, we consider readily available 2D materials, monolayered tungsten telluride ($\mathrm{WTe_2}$) and aligned graphene-hexagonal boron nitride (gr-hBN) in our analysis to facilitate experiments. 
	
	Quantum wavefunctions of electrons in a crystal are Bloch functions, labelled by Bloch vector $\mathit{\bf{k}}$ (quantum number), which gives the crystal momentum $\mathit{\hbar\bf{k}}$ that is conserved due to translational symmetry of the crystalline lattice. Bloch vector belongs to a unit cell of reciprocal space lattice, and Bloch functions are periodic with respect to $\mathit{\bf{k}}$. Adiabatic evolution of a Bloch function with $\mathit{\bf{k}}$ varying across a periodic unit cell is thus a cyclic one, picking up a geometric phase given by $\mathit{\gamma_\alpha=i\displaystyle{\oint{dk \left<u\!~(\bf{k})\middle|\frac{\partial}{\partial k_\alpha}\middle|u\!~(\bf{k})\right>}}}$, $\alpha$ denotes the direction in momentum space and $\mathit{u\!~(\bf{k})}$ the cell-periodic wavefunction of its state at $\mathit{\bf{k}}$ (we do not include band index for simplicity). The integrand is the Berry connection $\mathit{A_\alpha\bf(k)}$,which is analogous to vector potential in electromagnetism, and the loop integral can also be determined as a surface integral of the Berry curvature  $\mathit{\Omega_\delta{\!~(\bf{k})}\equiv\displaystyle{\varepsilon_{\delta\beta\alpha} \frac{\partial}{\partial k_\beta}\it{A_\alpha\bf{\!~(k)}}}}$, which is analogous to magnetic field governing the trajectory of electronic motion in reciprocal space \cite{berry1984quantal}. $\mathit{\Omega_\delta\bf{\!~(k)}}$, the curvature of quantum geometry of electrons is non-zero only along $\mathit{\hat{z}}$ direction in 2D ($\mathit{\delta=z}$) and vanishes identically in a crystal with time reversal and inversion symmetries.
	
\begin{figure}
		\includegraphics[width=8.6cm,keepaspectratio]{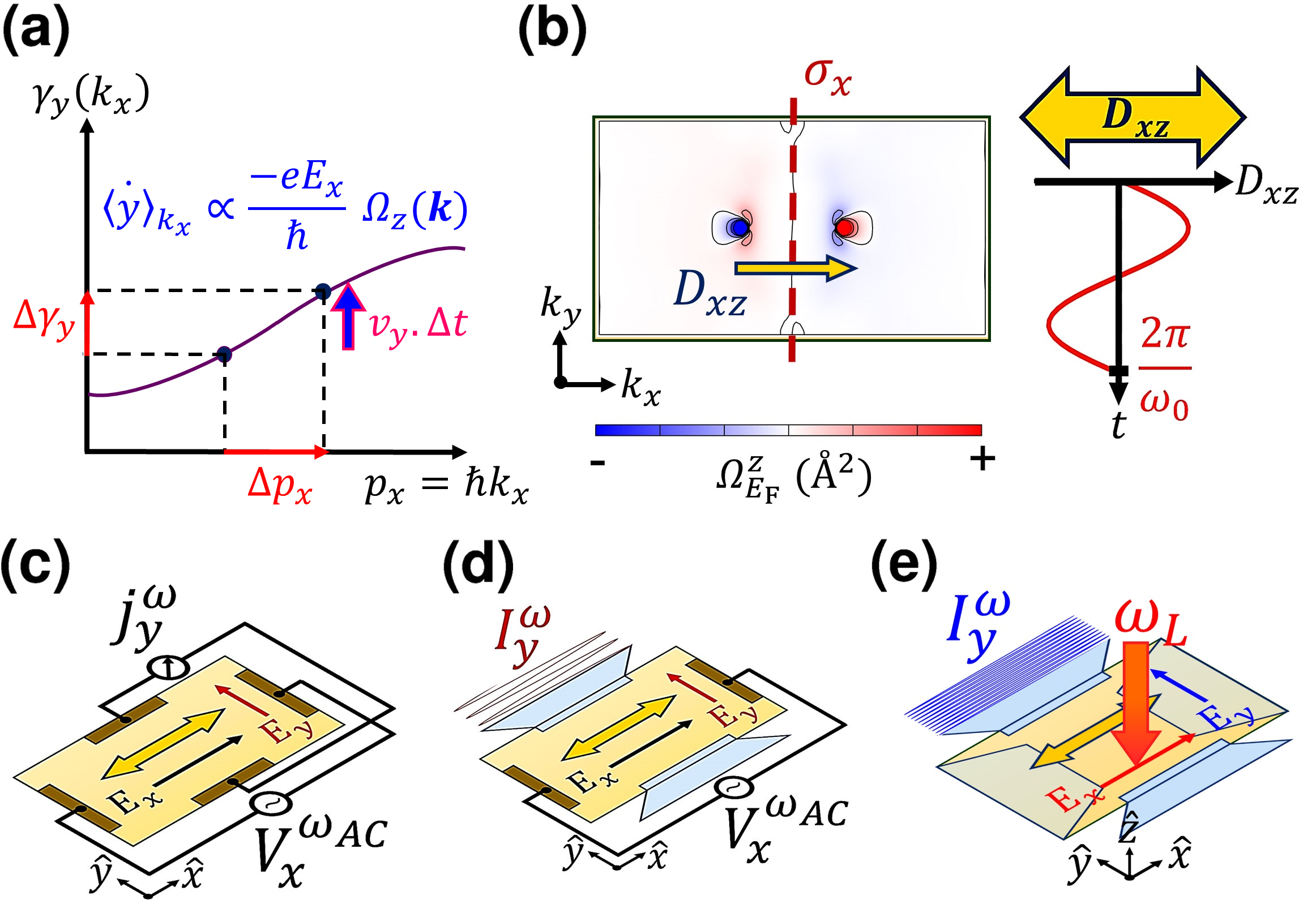}
		\caption{\label{tvel} \textbf{Quantum geometric origin of the anomalous Hall velocity and vibrational spectroscopy.} Transverse velocity $\mathit{\left<\dot{y}\right>_{k_x}}$ of an electron (a) driven by a force $\mathit{F_x=-eE_x}$ along $\mathit{\hat{k_x}}$ direction is associated with the change in its $\mathit{y}$-coordinate or Berry phase $\mathit{\gamma_y}$ generated by the Berry curvature $\mathit{\Omega_z\!~(\bf{k})}$, a measure of quantum geometric curvature. Structure with a single mirror symmetry $\mathit{\sigma_x}$ (b) permits $\mathit{{T_d}-\mathrm{WTe_2}}$ monolayer to exhibit nontrivial $\mathit{\Omega_z\!~(\bf{k})}$ shown as a contour-plot, and Berry curvature dipole $\mathit{D_{xz}}$ related to $\mathit{\displaystyle{\frac{\partial}{\partial k_x} \Omega_z\!~(\bf{k})}}$. A GQuES-active vibrational mode at frequency $\mathit{\omega_\mathrm{0}}$ and amplitude $\mathit{u_\mathrm{0}}$ dynamically lowers the symmetry of crystal, permitting oscillations in $\mathit{D_{xz}}$ (highlighted by double-headed yellow arrow in the right sub-panel of b). A schematic of an electronic device (c) proposed for GQuES based transport spectroscopy measures the transverse current $\mathit{j_y^\omega}$ in response to an applied voltage $\mathit{V_x^{\omega_{AC}}}$ at low frequency ~($\mathit{\omega_{AC} \leqslant \omega_\mathrm{0} <}$ 10 GHz), originating from the oscillations in $\mathit{D_{xz}}$ of a 2D crystal. A schematic of device (d) emits radiation $\mathit{I_y^\omega}$ polarized with electric field $\mathit{E_y}$ in response to an applied voltage $\mathit{V_x^{\omega_{AC}}}$ when vibrational frequencies are in THz. A schematic of device (e) proposed for GQuES based THz/optical spectroscopy to measure intensity $\mathit{I_y^\omega}$ of radiation polarized with electric field $\mathit{E_y}$ emitted in response to an incident light of frequency $\mathit{\omega_L \geqslant \omega_\mathrm{0}}$ polarized along $\mathit{\hat{x}}$ direction.}
	\end{figure}

	However, $\mathit{\Omega_z\!~(\bf{k})}$ can be nonzero in low symmetry 2D crystals and has an interesting effect on electronic motion: it gives rise to anomalous velocity in direction perpendicular to the motion driven by an external force. In the semiclassical treatment, force $\mathit{F_x=-eE_x}$, due to an external electric field $\mathit{E_x}$, changes the crystal momentum ($\mathit{\hbar k_x}$) by $\mathit{\hbar\Delta k_x=F_x \Delta t}$. As $\mathit{\gamma_y}$ along a periodic path in $\mathit{\hat{y}}$-direction gives the spatial position of electron, the change in $\mathit{k_x}$ due to $\mathit{F_x}$ results in an electronic drift along $\mathit{\hat{y}}$-direction when $\mathit{\displaystyle{\frac{\partial}{\partial \mathit{k_x}}\it{A_y\!~(\bf{k})}}\sim \Omega_z\!~(\bf{k})}$ is nonzero, termed as anomalous velocity [see Fig. \ref{tvel} (a)]. This transverse velocity is relevant to the anomalous Hall effect even in the absence of external magnetic field ($\mathit{B}$=0). The off-diagonal component of the Hall conductivity is $\mathit{\sigma_{xy}=-\displaystyle{\frac{e^\mathrm{2}}{\hbar} \iint{\frac{d^\mathrm{2}\bf{k}}{~(\mathrm{2}\pi)^\mathrm{2}}\Omega_z\!~(\bf{k})\ \mathit{f\!~(\varepsilon_{\bf{k}})}}}}$, where $\mathit{f\!~(\varepsilon_{\bf{k}})}$ is the Fermi-Dirac distribution function \cite{thouless1982quantized}.
	
	Even though low symmetry of a 2D crystal may allow a nonzero Berry curvature, the time reversal symmetry (TRS) requires $\mathit{\Omega_z\!~(\bf{k})=-\mathit{\Omega_z\!~(\bf{-k})}}$, giving a vanishing Hall conductivity $\mathit{\sigma_{xy}}$. Consequently, the \textit{linear} anomalous Hall effect is not possible in most non-magnetic systems. However, an applied field $\mathit{E_x}$ shifts the Fermi surface in k-space, giving an \textit{asymmetric} non-equilibrium distribution $\mathit{f}$ of electrons (see Supplemental Material \cite{SM,moore2010confinement}). As a result, \textit{nonlinear} Hall current depends quadratically \cite{sodemann2015quantum} on the applied field $\mathit{E_x}$ and is determined by the Berry curvature dipole $\mathit{\displaystyle{D\colon j_y \propto E_x^\mathrm{2} \cdot D}}$, where
	\begin{subequations}
	\begin{align}
		D&=\displaystyle{\iint_{\bf{k}} \left[\frac{\partial}{\partial{k_x}}\ \mathit{\Omega_z\!~(\bf{k})} \right] f(\varepsilon_{\bf{k}})} \\
		&\displaystyle{ =-\frac{\mathrm{1}}{\hbar}\iint_{\bf{k}} \left(\mathit{v_x\!~(\bf{k})}\right) \left[ \frac{\partial}{\partial \varepsilon_{\bf{k}}} \mathit{f}\!~(\varepsilon_{\bf{k}}) \right] \mathit{\Omega_z\!~(\bf{k})}}.
	\end{align}
	\end{subequations}
	For an applied AC field of frequency $\mathit{\omega_{AC}}$, the second-order NLH effect is observable in the forms of rectified Hall current $\mathit{j_y\!~(\mathrm{0})}$ and second-harmonic current $\mathit{j_y\!~(\mathrm{2}\omega_{AC})}$, when the Berry curvature dipole ($\mathit{D}$) (see Supplemental Material \cite{SM}) is nonzero \cite{sodemann2015quantum,sinha2022berry}. 
	
	$\mathit{D}$ is a tensor in 3D and a vector in 2D subject to significant significant symmetry constraints. Non-centrosymmetric Weyl semi-metals (WSM) are likely good candidates for observing NLH effect, as a consequence of large $\mathit{\Omega_z\!~(\bf{k})}$ arising near the Weyl points that act as magnetic monopoles \cite{zhang2018berry,zhang2022giant}. Second order NLH effect was first observed in $\mathrm{WTe_2}$ bilayer \cite{ma2019observation} and electrically switchable $\mathit{D}$ was demonstrated in $\mathit{T_d}\mathrm{-WTe_2}$ monolayer (type-II WSM) \cite{xu2018electrically}. Theoretical prediction of the Berry-connection polarizability (BCP) tensor \cite{gao2014field} was validated later in an experiment \cite{lai2021third} probing BCP induced third-order NLH effect in thick $\mathit{T_d}\mathrm{-MoTe_2}$. 
	
	In a 2D crystal, nonzero $\mathit{\Omega_z\!~(\bf{k})}$ leads to finite $\mathit{D}$ ($\mathit{D_{bz}}$ with ($\mathit{b=x,y}$), as permitted by the symmetries of the crystal (see Supplemental Material \cite{SM}).  For example, consider a 2D crystal having only the $\mathit{\sigma_x}$ reflection symmetry, as shown in Fig. \ref{tvel} (b). Since $\mathit{\Omega_z\!~(\bf{k})}$ is odd under the TRS and $\mathit{\sigma_x\!~(k_x,k_y)\rightarrow~(-k_x,k_y)}$ symmetries, its $\mathit{D_{xz}}$ is nonzero and its magnitude is determined by $\mathit{\Omega_z\!~(\bf{k})}$ and the group velocity of electrons at the Fermi energy \cite{sodemann2015quantum}. On the other hand, $\mathit{D_{yz}}$ vanishes in such a crystal as the velocity along $\mathit{\hat{y}}$ direction is even under the reflection symmetry $\mathit{\sigma_x}$. A single, in-plane mirror symmetry ($\mathit{\sigma_x}$ or $\mathit{\sigma_y}$) of a 2D crystal permits its $\mathit{D}$, perpendicular to the mirror line ($\mathit{D_{xz}}$ or $\mathit{D_{yz}}$) to be nonzero [see Fig. 1(b)]. 
	
	In this work, we consider dynamically broken symmetry states with nonzero $\mathit{D}$ and their observable signatures in the NLH effect, with focus on vibrational excitations in 2D materials. Symmetry of a structure can get lowered with distortion during a normal mode of vibration, allowing nonzero oscillations in the $\mathit{D}$. This naturally gives rise to an observable NLH response modulated with the vibrational frequency $\mathit{\omega_{\mathrm{0}}}$ [Figs. \ref{tvel}(c) – (e)]. In response to an electric field $\mathit{E_x(t)=\displaystyle{Re~({\mathcal{E}_x \cdot e^{i\omega_{AC}t}})}}$ applied along $\mathit{\hat{x}}$ direction of a 2D crystal, symmetry-allowed $\mathit{D}$ generates a NLH current at twice the applied frequency $\mathit{j_y~(\mathrm{2}\omega_{AC})}$ and at zero frequency $\mathit{j_y~(\mathrm{0})}$ \cite{sodemann2015quantum}. For a dynamical structural distortion with amplitude ($\mathit{u}$) induced by a vibrational mode of frequency $\mathit{\omega_\mathrm{0} > \mathrm{2}\omega_{AC}}$, the NLH current is expected at frequencies $\mathit{\omega_\mathrm{0}}$ and $\mathit{\omega_\mathrm{0} \pm \mathrm{2}\omega_{AC}}$, as a consequence of induced $\mathit{D}$. Thus, $\mathit{\partial D/\partial u \neq \mathrm{0}}$ defines the selection rule for GQuES spectroscopy.
	
	Expanding up to first order in the amplitude $\mathit{u}$ of the vibrational mode, $\mathit{\displaystyle{D=D^{\!~(\mathrm{0})}+u\left.\frac{\partial D}{\partial u}\right|_{u=\mathrm{0}}}}$ with $\mathit{u=u_\mathrm{0} cos\!~(\omega_\mathrm{0} t)}$. For only $\mathit{q=\mathrm{0}}$ ($\mathit{\lambda \rightarrow \infty}$) phonon, $\mathit{\displaystyle{\frac{\partial D}{\partial u}\neq\mathrm{0}}}$, as a consequence of the conservation of momentum \cite{loudon1963theory}, and $\mathit{\displaystyle{\frac{\partial D}{\partial u_{\roarrow{\bf{q}}\neq\mathrm{0}}}=\mathrm{0}}}$. Within relaxation time approximation, frequency-dependent NLH current is 
	
	\begin{multline}
			j_y\!~(\omega)=-\frac{\epsilon_{abc} e^3 \tau}{4~(1+i\omega \tau)} \mathcal{E}_x^2 \Big\{2D\Big[\delta\!~(\omega - 2\omega_{AC})+ \delta\!~(\omega) \Big] \\ + u_0 \left.\frac{\partial D}{\partial u}\right|_{u=0} \Big[\delta\!~(\omega - ~(\omega_0+2\omega_{AC}))+\delta\!~(\omega-\vert\omega_0 - 2\omega_{AC}\vert) \\+2\delta\!~(\omega-\omega_0)\Big] \Big\}  
	\end{multline}
	
	where $\mathit{e}$ and $\mathit{\tau}$ are the electronic charge and relaxation time respectively. For a 2D crystal with an inherent $\mathit{D}$, both the terms contribute to NLH current: first one at (i) $\mathit{\omega=\mathrm{0}}$ (rectification), (ii) $\mathit{\mathrm{2}\omega_{AC}}$ (second harmonic generation), which have been observed \cite{ma2019observation}, and the second one, \textit{yet to be observed}, at $\mathit{\omega=}$ (iii) $\mathit{\omega_\mathrm{0}}$ and (iv) $\mathit{\left|\omega_\mathrm{0}+\mathrm{2}\omega_{AC}\right|}$. When the symmetries of a 2D crystal force its intrinsic $\mathit{D}$ to vanish, only the second term containing $\mathit{\partial D/\partial u}$ is responsible for the NLH current at frequencies $\mathit{\omega_\mathrm{0}}$ and $\mathit{\left|\omega_\mathrm{0} \pm \mathrm{2}\omega_{AC}\right|}$. It can be measured as a Hall voltage (Fig. \ref{tvel} (c)), which is feasible for low frequency vibrations (example, acoustic phonons). As $\mathit{\omega_\mathrm{0}}$ (optical phonons) is typically in THz, its transport measurement in a circuit is challenging [Fig. \ref{tvel} (c)], and may rather be detected as THz radiation with polarization $\mathit{E_y}$ emitted by the sample aided by a transverse antenna structure [Fig. \ref{tvel} (d)]. GQuES based optical (or EM waves) spectroscopy is also possible with LASERs [Fig. \ref{tvel} (e)], which involves normal incidence of THz radiation or light with $\mathit{E_x}$ polarization at frequency $\mathit{\omega_L}$ and detection of emitted electromagnetic radiation with $\mathit{E_y}$ polarization at frequencies $\mathit{\mathrm{2}\omega_L \pm \omega_\mathrm{0}}$. 
	
	We now present material-specific evidence to support these ideas using first-principles theoretical calculations of (i) vibrationally induced $\mathit{D}$ of the centrosymmetric $\mathit{T\sp{\prime}}$$\mathrm{-WTe_2}$ monolayer and (ii) spontaneous as well as vibrationally induced $\mathit{D}$ of non-centrosymmetric $\mathit{T_d}$$\mathrm{-WTe_2}$ monolayer. $\mathit{T\sp{\prime}}$$\mathrm{-WTe_2}$ monolayer belongs to \textit{P21/m} space group with $\mathit{C_{2h}}$($\mathit{E, C_{2x}, \hat{I}, \sigma_x}$) point group symmetry, while $\mathit{T_d}$$\mathrm{-WTe_2}$ monolayer belongs to \textit{P1m1} space group with $\mathit{C_s}$($\mathit{E, \sigma_x}$) point group symmetry. $\mathit{T\sp{\prime}}$$\mathrm{-WTe_2}$ and $\mathit{T_d}$$\mathrm{-WTe_2}$ monolayers are time-reversal symmetric, type-II WSM with electron and hole pockets crossing the Fermi energy ($\mathit{E_\mathrm{F}}$) \cite{soluyanov2015type}. 
	
\begin{figure}
		\includegraphics[width=8.6cm,keepaspectratio]{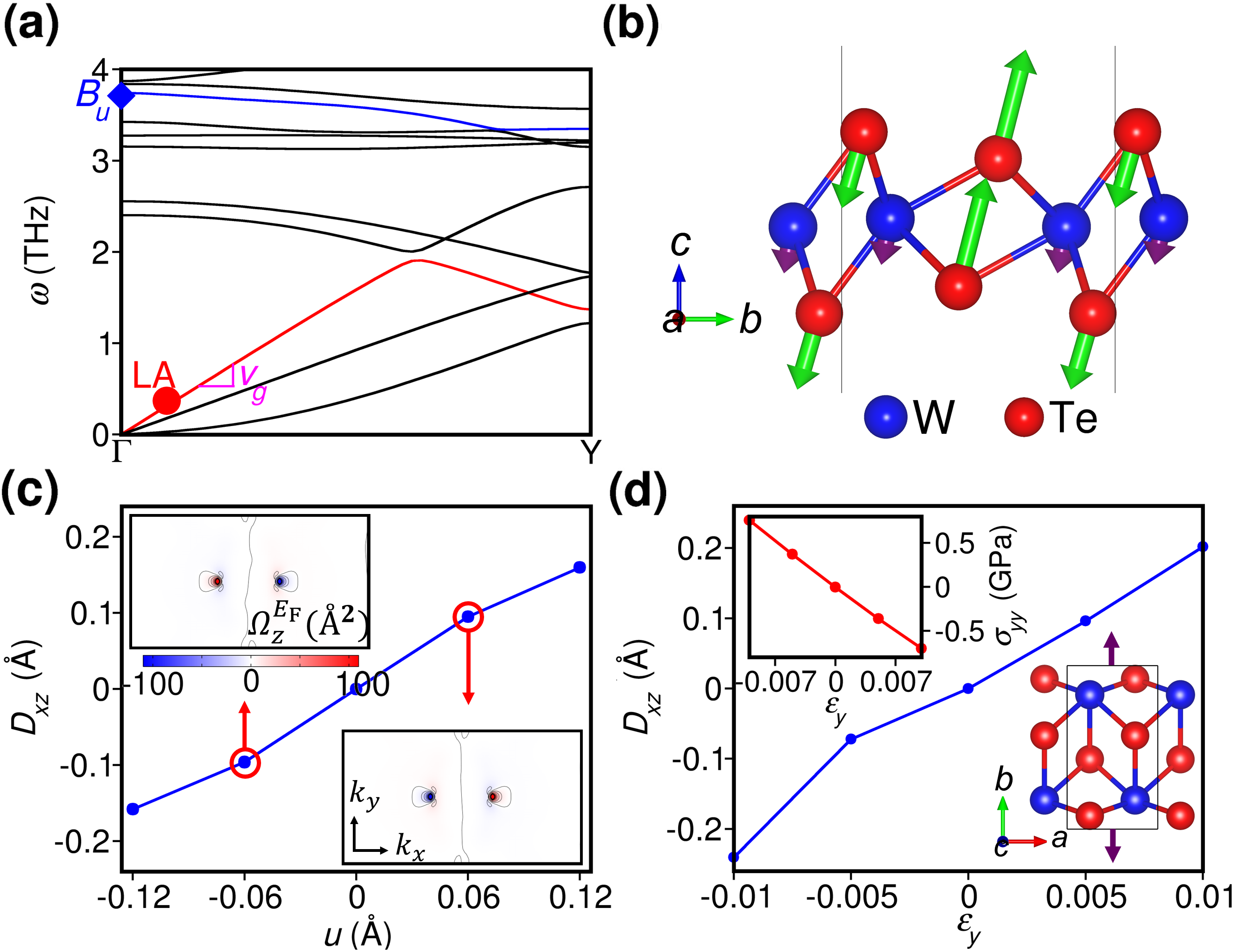}
		\caption{\label{wte2} \textbf{GQuES-activity of phonons of $\mathrm{\bf{WTe_2}}$ monolayer.} Phonon dispersion (a) of centrosymmetric $\mathit{T\sp{\prime}}$$\mathrm{-WTe_2}$ monolayer (ML) with GQuES-active optic phonon $\mathit{B_u}$ at $\mathit{\omega_\mathrm{0}}$=3.74 THz (blue diamond) and longitudinal acoustic (LA) mode at $\mathit{\omega_\mathrm{0}}=\!~(\frac{\mathrm{2}\pi}{L} v_g)$=3.17 GHz (red circle) that constitutes a natural vibration of a sample of lateral size $\mathit{L}$=4 $\mathit{\mu}$m. These vibrations dynamically lower the structural symmetry and induce oscillations in $\mathit{D}$. Structural distortion induced by collective atomic displacements $\mathit{~(u)}$ of the $\mathit{B_u}$ normal mode (b) and induced variation in $\mathit{D_{xz}}$ that is linear with amplitude $\mathit{u}$ (c). Insets in c show contour plots of $\mathit{\Omega_z\!~(\bf{k})}$ of $\mathit{T\sp{\prime}}$$\mathrm{-WTe_2}$ ML at Fermi energy $\mathit{~(E_\mathrm{F})}$ exhibiting reversal in the polarity of Berry curvature dipole when the structure is distorted with $\mathit{u<\mathrm{0}}$ and $\mathit{u>\mathrm{0}}$ (indicated by red arrows). Application of uniaxial strain along $\mathit{\hat{b}}$ direction of $\mathit{T\sp{\prime}}$$\mathrm{-WTe_2}$ ML ($\mathit{\varepsilon_y}$) preserves only its $\mathit{\sigma_x}$ symmetry and induces $\mathit{D_{xz}}$ that varies linearly with respect to $\mathit{\varepsilon_y}$ (d). Insets in (d) show linear variation in mechanical stress $\mathit{\sigma_{yy}}$ (GPa) with respect to $\mathit{\varepsilon_y}$ (violet arrows). The slopes of these quasi-linear curves in (c) and (d) determine the GQuES activity of a phonon.} 
	\end{figure}

	Inversion symmetry of $\mathit{T\sp{\prime}}$$\mathrm{-WTe_2}$ monolayer forces its intrinsic Berry curvature (and hence $\mathit{D}$) to vanish, while the distortions induced by a vibrational mode, for example $\mathit{B_u}$ at $\mathit{\omega_\mathrm{0}}$=3.7 THz [Figs. \ref{wte2} (a) and (b)], lowers its symmetry to a single mirror plane $\mathit{\sigma_x}$ and permits nonzero $\mathit{D_{xz}}$, induced orthogonal to the $\mathit{\sigma_x}$ mirror line [Fig. \ref{wte2}(c)]. Sizeable $\mathit{\Omega_z\!~(\bf{k})}$, of the order of few 100 $\mathrm{\AA^2}$, concentrated near the Weyl points (WP), oscillates in its polarity as the $\mathit{T\sp{\prime}}$$\mathrm{-WTe_2}$ structure vibrates with $\mathit{B_u}$ mode [see insets of Fig. \ref{wte2} (c)]. Though the induced $\mathit{D}$ at $\mathit{E_\mathrm{F}}$ is of the order of 10$\mathrm{^{-1}}$ \AA, it is the sensitivity of $\mathit{D}$ with respect to the amplitude of a specific vibrational mode that is relevant to GQuES. Noting that $\mathit{B_u}$ is a polar phonon and can be selectively excited with IR radiation at a frequency of $\mathrm{\sim}$ 3.7 THz, we expect its GQuES Hall signal to be enhanced systematically with the IR-pump, though it is not necessary. This is somewhat similar to the metastable topological phases of HgTe obtained through coherent excitation of IR-active phonon mode \cite{shin2024light}.
	
	A long wave-length ($\mathit{q\rightarrow\mathrm{0}}$) longitudinal acoustic (LA) phonon [Fig. \ref{wte2} (a)] is a quantized form of a strain wave, which can induce oscillations in $\mathit{D}$ if the associated strain lowers the lattice symmetry, for example uniaxial strain along \textit{y}-direction [Fig. \ref{wte2} (d)]. A natural mode of vibration along $\mathit{\hat{b}}$ of the $\mathit{T\sp{\prime}}$$\mathrm{-WTe_2}$ monolayer, shown in the inset of Fig. \ref{wte2} (d), induces $\mathit{D}$ and exhibits GQuES activity. Similarly, $\mathit{A\sp{\prime}}$ and $\mathit{A^{\prime\prime}}$ phonons of $\mathit{T_d}$$\mathrm{-WTe_2}$ monolayer are GQuES active (see Supplemental Material \cite{SM} and would generate NLH responses at frequencies $\mathit{\omega_\mathrm{0}}$ and $\mathit{\left|\omega_\mathrm{0} \pm \mathrm{2}\omega_{AC}\right|}$ (in addition to the NLH responses at frequencies of $\mathit{\mathrm{2}\omega_{AC}}$ and $\mathit{\omega}$=$\mathrm{0}$).
	
\begin{figure}
		\includegraphics[width=8.6cm,keepaspectratio]{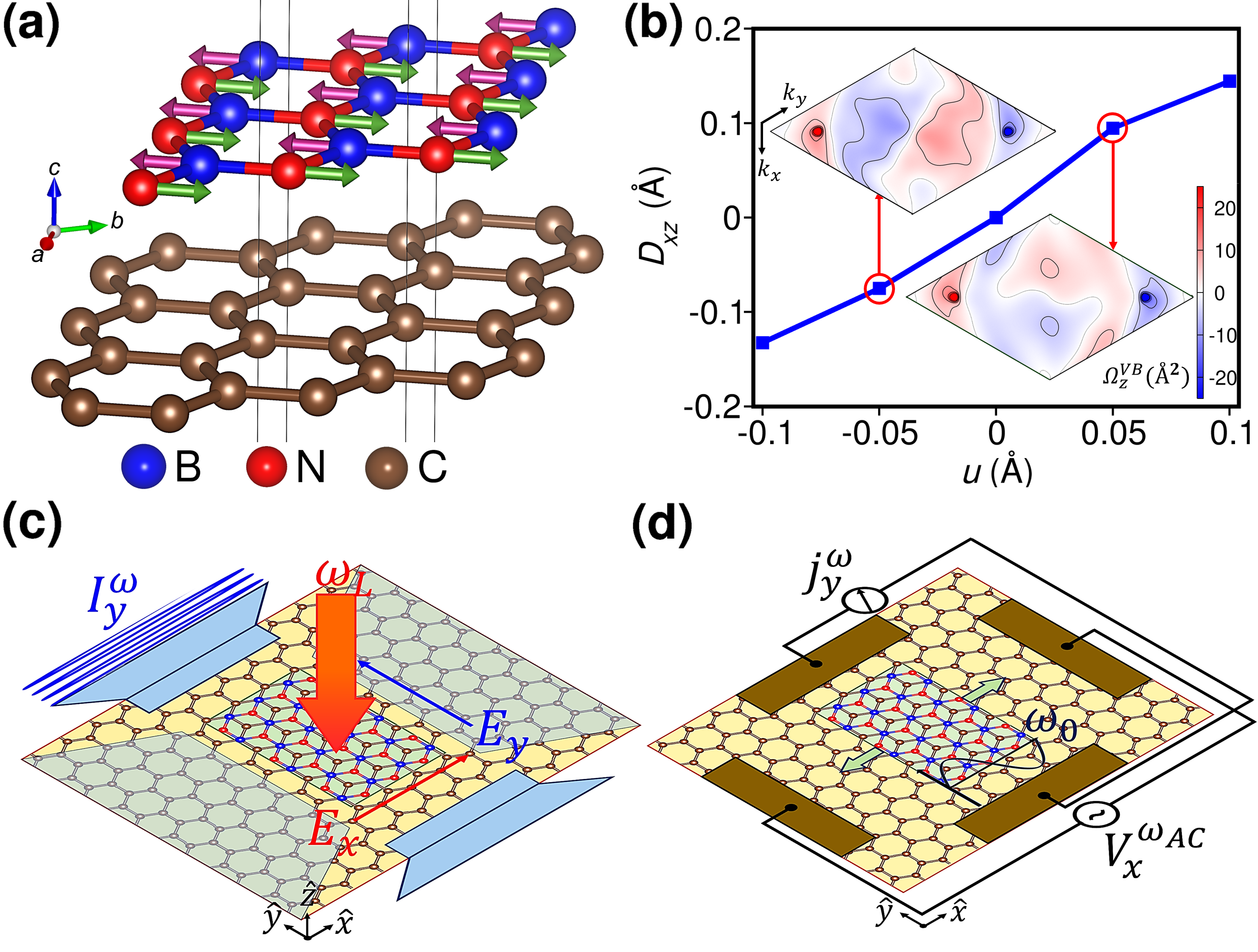}
		\caption{\label{grhbn} \textbf{Substrate induced GQuES activity of phonons in aligned gr-hBN.} While the 3-fold rotational symmetry of aligned gr-hBN forces its $\mathit{D}$ to vanish, its GQuES-active vibrational mode $\mathit{E}$ (doubly degenerate at $\mathit{\omega_\mathrm{0}}$=41.9 THz), involving atomic vibrations shown in (a) lowers the structural symmetry and induces oscillating $\mathit{D_{xz} \propto D\sp{\prime} u \cos(\omega_\mathrm{0} t)}$, where $\mathit{D\sp{\prime}}$ is given by the slope of variation in $\mathit{D_{xz}}$ with $\mathit{u}$, the amplitude of $\mathit{E}$ mode (b). Insets in (b) are contour plots of $\mathit{\Omega_z\!~(\bf{k})}$ of aligned gr-hBN at valence band edge distorted with $\mathit{u<\mathrm{0}}$ and $\mathit{u>\mathrm{0}}$, whose asymmetric distribution generates $\mathit{D}$. $\mathit{E}$ phonon distort hBN ML along cartesian $\mathit{\hat{y}}$ (a) and $\mathit{\hat{x}}$ directions (see Supplemental Material \cite{SM}), which break $\mathit{C_{3z}}$ and $\mathit{C_{3z}+\sigma_x}$ symmetries, thereby allowing nonzero $\mathit{D_{xz}}$ (b) and $\mathit{D_{bz} ~(b=x,y)}$ respectively. Schematics of experimental setups for observation of emission of THz radiation $\mathit{I_y^\omega}$ as the nonlinear Hall response to incident LASER at 2.52 THz (red beam) (c) or as Hall current in circuit mode (d). While a vibration is localized on h-BN (shown with arrows in (d)), the NLH response is generated by the electronic states near Dirac points of graphene (a substrate).}
	\end{figure}

	For GQuES spectroscopic signatures to be observable in a 2D crystal, (a) condition of low crystal symmetry needs to be satisfied and (b) quantum geometry of electronic bands needs to host non-zero $\mathit{\Omega_z\!~(\bf{k})}$ at accessible energy levels (i.e., at energies in the vicinity of chemical potential). Not all materials satisfying the symmetry conditions exhibit nontrivial $\mathit{\Omega_z\!~(\bf{k})}$. For example, h-Boron nitride (h-BN) is a wide-gap insulator, commonly used in devices based on two-dimensional materials \cite{caldwell2019photonics}. A large band gap weakens its conductivity and Hall responses, even if nonzero $\mathit{D}$ may be permitted by the dynamically lowered symmetries of h-BN crystal. These limitations, we show, can be overcome by aligning it on graphene. A monolayer of graphene aligned on top of h-BN in the AB-stacking (aligned gr-hBN, see Fig. \ref{grhbn} (a)) exhibits a narrow band gap of 45 meV at the Dirac points due to the interlayer crystal field. While the point group symmetry $\mathit{C_{3v}}$($\mathit{E, C_{3z},\sigma_x}$) of aligned gr-hBN permits its $\mathit{\Omega_z\!~(\bf{k})}$ to be nonzero, its intrinsic $\mathit{D}$ vanishes due to the three-fold rotational symmetry. By stacking a system with symmetry allowed GQuES-active modes (eg. h-BN) on a two-dimensional graphene with nontrivial $\mathit{\Omega_z\!~(\bf{k})}$, the second criterion necessary to observe GQuES signatures of a crystal is satisfied.
	
	A structural distortion induced by a doubly degenerate vibrational mode $\mathit{E}$ at $\mathit{\omega_\mathrm{0}}$=41.9 THz [Fig. \ref{grhbn} (a)] of h-BN breaks the $\mathit{C_{3z}}$ symmetry and permits a nonzero $\mathit{D}$, as shown in Fig. \ref{grhbn} (b). This GQuES active vibrational mode induces symmetry lowering distortions primarily in h-BN and its interaction with graphene modulates the electronic structure at the Dirac points of graphene inducing oscillations in $\mathit{D}$ [(Fig. \ref{grhbn} (b)]. Activity of $\mathit{E}$ mode of h-BN can be thus observed in the GQuES spectrum using LASER as a probe in the form of $\mathit{\omega}$-dependent $\mathit{I_y^\omega}$ emitted by the supporting antenna (Fig. \ref{grhbn} (c)) and using EM waves as a probe, as shown in Fig. \ref{grhbn} (d). Similarly, macroscopic strain vibration of aligned gr-hBN corresponding to an LA phonon also lowers the crystal symmetry inducing oscillations in $\mathit{D}$ and is predicted to be observable in nonlinear Hall transport spectrum [Fig. \ref{grhbn} (d)].
	
	Using results in Figs. \ref{wte2} and \ref{grhbn}, we now present system-specific GQuES spectra that may be validated with experimental observations in two modes: (a) transport mode, when the vibrational frequency of the sample $\mathit{\omega_\mathrm{0}\!}<$10 GHz and the frequency of probing field $\mathit{\omega_{AC}<\omega_\mathrm{0}/\mathrm{2}}$, and the signal is read as either Hall current [Fig. \ref{tvel} (c)] or emitted THz radiation [Fig. \ref{tvel} (d)], and (b) EM THz/optical mode, when $\mathit{\omega_\mathrm{0}<\mathrm{2}\omega_L}$, $\mathit{\omega_L}$ being the frequency of probing EM field imposed with a LASER [Fig. \ref{tvel} (e)], and signal is in the frequency dependent EM waves with transverse polarization emitted by the sample surrounded by antennas [Figs. \ref{tvel} (d) and (e)]. While the transport mode requires electrical contacts for measurements, the THz/optical mode is contact-less.
	
\begin{figure}
		\includegraphics[width=8.6cm,keepaspectratio]{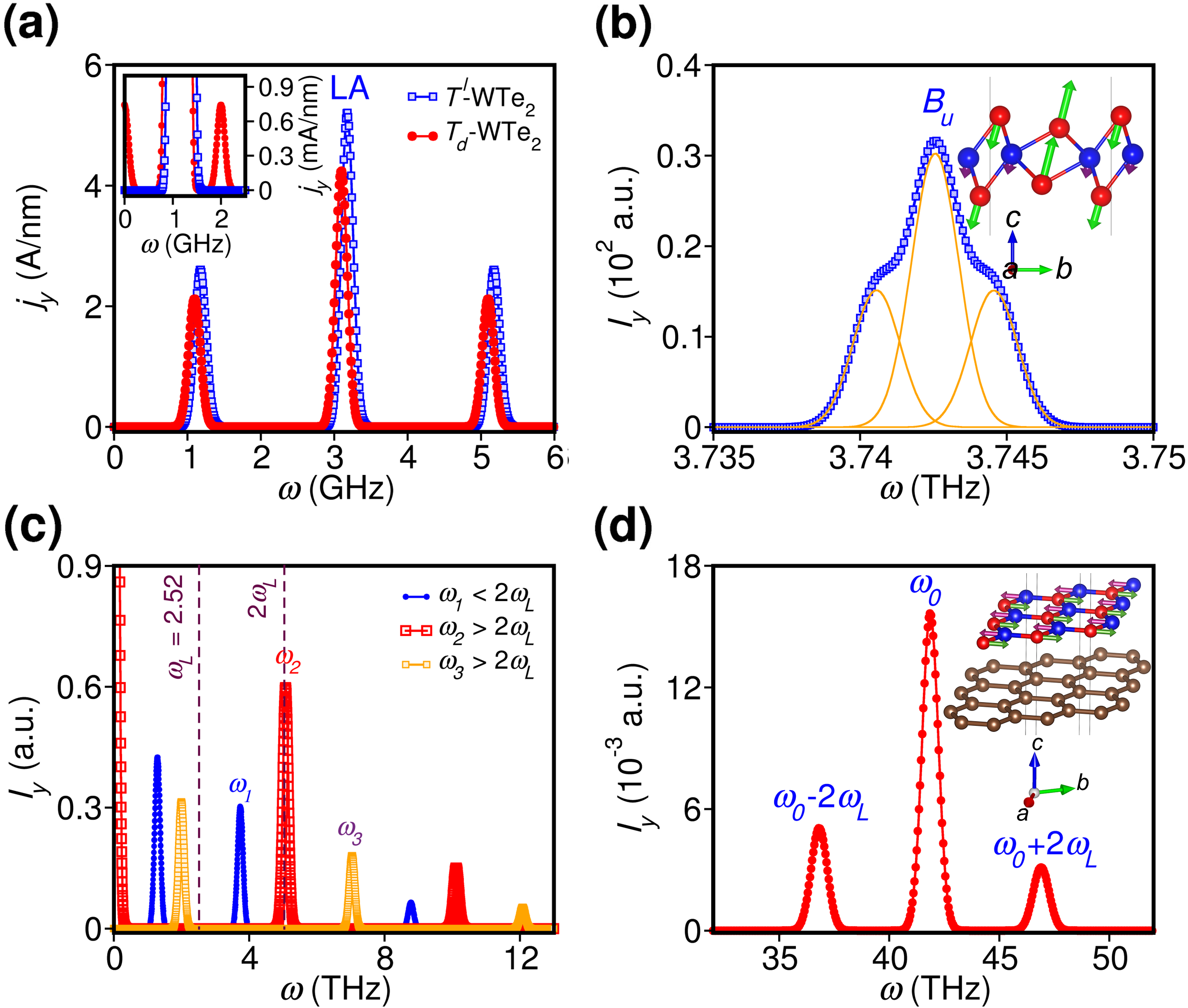}
		\caption{\label{tptint} \textbf{Predictions for experimental observations of GQuES spectra of 2D materials.} GQuES vibrational spectra from measurement of NLH current $\mathit{j_y~(\omega)}$ (a) and THz emission (b) in response to voltage $\mathit{V_x}$ ($\mathit{\omega_{AC}}$= 1 GHz) applied to $\mathrm{WTe_2}$. Intensity of the emitted THz radiation polarized along \textit{y}-direction in response to LASER ($\mathit{\omega_L}$= 2.52 THz) with \textit{E}-polarization along \textit{x}-axis incident on $\mathrm{WTe_2}$ (c) and aligned gr-hBN (d). Macroscopic strain mode of vibration of centrosymmetric $\mathit{T\sp{\prime}\mathrm{-WTe_2}}$ (blue squares) and non-centrosymmetric $\mathit{T_d\mathrm{-WTe_2}}$ (red circles) monolayered samples of lateral size L=4 $\mathit{\mu}$m, corresponding to LA phonons at $\mathit{\omega_\mathrm{0}}$=3.17 GHz and $\mathit{\omega_\mathrm{0}}$=3.09 GHz respectively, manifest as 3 peaks each (a) at $\mathit{\omega_\mathrm{0}}$ and $\mathit{\omega_\mathrm{0} \pm \mathrm{2}\omega_{AC}}$. Additional peaks at $\mathit{\omega=\mathrm{0}}$ and $\mathit{\mathrm{2}\omega_{AC}}$ are evident in NLH current of $\mathit{T_d\mathrm{-WTe_2}}$ due to its intrinsic $\mathit{D}$ allowed by the broken inversion symmetry (inset of a). GQuES-active $\mathit{B_u}$ mode (inset) of $\mathit{T\sp{\prime}\mathrm{-WTe_2}}$ ML manifests as peaks in the intensity of emitted radiation (b) at $\mathit{\omega_0}$=3.74 THz and $\mathit{\omega=\omega_\mathrm{0} \pm \mathrm{2}\omega_{AC}}$ ($\mathit{\omega_{AC}}$=1 GHz). In response to LASER at frequency $\mathit{\omega_L}$=2.52 THz ($\mathit{\lambda_L=}$118.9 $\mathit{\mu}$m), GQuES-active phonons of $\mathit{T\sp{\prime}\mathrm{-WTe_2}}$ monolayer ($\mathit{\omega_\mathrm{1}}$=3.74 THz, $\mathit{\omega_\mathrm{2}}$=5.08 THz, $\mathit{\omega_\mathrm{3}}$=7.04 THz) are evident in peaks (c) at $\mathit{\omega=\mathrm{2}\omega_L \pm \omega_\mathrm{0}}$ (blue circles) and at $\mathit{\omega=\omega_\mathrm{0} \pm \mathrm{2}\omega_L}$ (red and orange squares) in the emitted radiation. Similarly, $\mathit{E}$ mode of aligned gr-hBN at $\mathit{\omega_\mathrm{0}}$=41.9 THz manifests as peaks in the GQuES spectrum (d) at $\mathit{\omega=\omega_\mathrm{0} \pm \mathrm{2}\omega_L}$ and $\mathit{\omega=\omega_\mathrm{0}}$ in the emitted radiation.}
	\end{figure}

	In the transport mode [Figs. \ref{tvel} (c) and (d)], we present Hall responses (see details in Supplemental Material \cite{SM}) of $\mathit{T\sp{\prime}}$ and $\mathit{T_d}$ structures of $\mathrm{WTe_2}$ to AC voltage applied at 1 GHz along the (\textit{x}-) direction (along its $\mathit{D}$). Frequencies of a macroscopic strain vibration of a sample [see the inset labelled as LA in Fig. \ref{tptint} (a)] of a few $\mathit{\mu}$m size (L) is in the range of a few GHz and can be measured as peaks in frequency ($\mathit{\omega}$) dependent Hall voltage at $\mathit{\omega=\omega_\mathrm{0}}$ and $\mathit{\omega_\mathrm{0} \pm \mathrm{2}\omega_{AC}}$. Precise frequency of the natural vibration is that of LA mode at $\mathit{q=\frac{\pi}{L}}$ [Fig. \ref{wte2}(a)]. In addition, we expect small peaks a $\mathit{\omega=\mathrm{0}, \mathrm{2}\omega_{AC}}$ for the non-centrosymmetric $\mathit{T_d}\mathrm{-WTe_2}$ due to its intrinsic $\mathit{D}$ [inset of Fig. \ref{tptint} (a)], which permit characterization of the inversion asymmetry of the structure. Similarly, the optical phonon $\mathit{B_u}$ of $\mathit{T\sp{\prime}}\mathrm{-WTe_2}$ manifests as three peaks (centred at its frequency $\mathit{\omega_\mathrm{0}}$=3.74 THz) in the intensity of emitted THz radiation [Fig. \ref{tptint} (b)] polarized along \textit{y}-direction. Clearly, the peak structure can be systematically controlled with the frequency $\mathit{\omega_{AC}}$ of the driving field.
	
	In the EM waves THz/optical mode of GQuES spectroscopy [Fig. \ref{tvel} (e)], we present responses of $\mathrm{WTe_2}$ [Fig. \ref{tptint} (c)] and aligned gr-hBN [Fig. \ref{tptint} (d)] to a LASER at $\mathit{\lambda_L}$=118.9 $\mathit{\mu}$m ($\mathit{\omega_L}$=2.52 THz) with electric field polarized along \textit{x}-direction. In the doubled frequency range, we expect emission of radiation with \textit{y}-polarization that peaks at several frequencies $\mathit{\omega=\omega_\nu}$ and $\mathrm{\left|5.04 THz\mathit{\pm \omega_\nu}\right|}$, where $\mathit{\omega_\nu}$ is the frequency of $\mathit{\nu^{\rm{th}}}$ GQuES-active mode. For $\mathit{T\sp{\prime}\mathrm{-WTe_2}}$, these peaks [Fig. \ref{tptint} (c)] correspond to different $\mathit{B_u}$ phonons [see Fig. \ref{tptint} (c)] with $\mathit{\omega_\nu}$=3.74, 5.08, 7.04 THz. In addition, peaks at $\mathit{\omega}$=5.04 THz and $\mathit{\omega}$=0 are expected in GQuES spectrum of $\mathit{T_d\mathrm{-WTe_2}}$ (see Supplemental Material \cite{SM}) due to its intrinsic $\mathit{D}$. Similarly, we predict that $\mathit{E}$ mode of aligned gr-hBN manifests itself as a peak at $\mathit{\omega_\mathrm{0}}$=41.9 THz along with side peaks at $\mathit{\omega}$=41.9$\mathit{\pm}$5.04 THz in the intensity of emitted IR radiation [Fig. \ref{tptint} (d)]. While Stokes and anti-Stokes peaks in a Raman spectrum are centred at $\mathit{\omega=\omega_L}$, the peaks manifested in the GQuES spectra are centred at $\mathit{\omega=\mathrm{2}\omega_L}$.
	
	The mechanism of GQuES based vibrational spectroscopy originates from the oscillations in the quantum geometric curvature of electronic structure, and can also be tested in graphene and $\mathrm{MoS_2}$ based 2D materials, where strain dependent $\mathit{D}$ has been reported \cite{sinha2022berry,battilomo2019berry,xiao2020two,son2019strain}. Our demonstration of the GQuES vibrational spectroscopy of aligned gr-hBN shows that a GQuES-active material could be used as a substrate in analysis of a system that is otherwise inactive. It could also be useful in translating signals in GHz or sub-GHz frequencies into those in the THz range [Figs. \ref{tptint}(b)], with a suitable choice of active material and modes of natural vibration. At higher frequencies $\mathit{\hbar\omega_L > E_g}$ (see Supplemental Material \cite{SM}), we expect quantum resonant processes to become relevant and make the GQuES spectra even richer, requiring quantum mechanical treatment. The GQuES vibrational spectroscopy demonstrated here with 2D crystals can be generalized to (a) other excitations that dynamically break crystal symmetry, and (b) 3D crystals, with manifestations of quantum geometry in even richer set of phenomena.
	
	 In summary, we have demonstrated from first-principles that \textit{lattice vibrations can be gauged using the oscillations they induce in the quantum geometry and Hall transport of electrons}. We expect it to translate into a powerful experimental technique to measure spectra of acoustic and optic phonons and other excitations.
	
	\begin{acknowledgments}
		RB acknowledges SASTRA Deemed University, Thanjavur for financial assistance through a fellowship. UVW acknowledges support from a JC Bose National Fellowship of SERB-DST, Government of India.
	\end{acknowledgments}
	
	The idea was conceived through discussion between RB and UVW. The calculations were performed by RB, a strong connection with experimental specifics was achieved through interaction between MMD, RB and UVW. The manuscript was written by all the three authors.
	
	\bibliography{GQuES}
	\nocite{*}
\end{document}


\title{Supplementary Materials: Sensing Vibrations using Quantum Geometry of Electrons}%
	\author{Bhuvaneswari R}%
	\email{bhuvana@jncasr.ac.in}
	\affiliation{Theoretical Sciences Unit, Jawaharlal Nehru Centre for Advanced Scientific Research, Bangalore, 560064, India.}
	\affiliation{School of Electrical \& Electronics Engineering, SASTRA Deemed University, Thanjavur, 613401, India.}
	\author{Mandar M Deshmukh}%
	\email{mandar.m.deshmukh@gmail.com}
	\affiliation{Department of Condensed Matter Physics and Materials Science, Tata Institute of Fundamental Research, Mumbai, 400005, India.}
	\author{Umesh V Waghmare}%
	\email{waghmare@jncasr.ac.in}
	\affiliation{Theoretical Sciences Unit, Jawaharlal Nehru Centre for Advanced Scientific Research, Bangalore, 560064, India.}
	\date{\today}%
	
	\maketitle
	\newpage
	
	\tableofcontents
	\listoftables
	\listoffigures
	\newpage
	
	\subsection{Computational Details}
		We performed comprehensive first-principles theoretical analysis of T’-WTe2 and Td-WTe2 monolayers and aligned gr-hBN in Quantum ESPRESSO package \cite{Giannozzi2009} with projector augmented wave (PAW) potential based on density functional theory (DFT). The exchange-correlation energy of electrons is treated with a generalized gradient approximation (GGA) and functional of Perdew-Burke-Ernzerhof (PBE) parametrized form \cite{Perdew1997}. The structures were fully optimized using conjugate gradient (CG) algorithm until the mismatch between the experimentally reported lattice parameters and the DFT calculated values are less than 2 \% and the Hellmann-Feynman forces on each atom were smaller than 0.01 eV/Å. A uniform mesh of 8 x 8 x 1 k-points was used to sample the integrations over the Brillouin zone and a vacuum padding of 16 Å was employed to prevent the periodic images from interacting along the layer direction ($\mathit{\hat{z}}$). Grimme-d2 \cite{Grimme2006} van der Waals corrections were used to account the interlayer interactions in the bilayer properly. Spin-orbit coupling (SOC) with fully-relativistic pseudopotentials were included in determining the electronic structures of $\mathit{T\sp{\prime}}\mathrm{-WTe_2}$, $\mathit{T_d}\mathrm{-WTe_2}$ monolayers and aligned gr-hBN.

		Phonon calculations were carried out with linear response method over a mesh of 4 x 4 x 1 q-points, which were Fourier interpolated to obtain phonons at arbitrary q-points to ensure the dynamical stability of the structures. The structures are then distorted with the eigen displacements of the GQuES-active vibrational mode at $\mathit{\Gamma}$-point ($\mathit{q}$=0) that lowers the relevant symmetries of the crystal. The DFT Bloch wavefunctions were projected to maximally localized Wannier functions (MLWFs) by WANNIER90 code \cite{Wang2006,Mostofi2014} with individual spread less than 5 $\mathrm{\AA^2}$. The tight-binding model constructed by WANNER90 after convergence of the wannierisation procedure was given as input to WannierTools \cite{Wu2018} and WannierBerri \cite{Tsirkin2021} package to calculate Berry curvature and Berry curvature dipole, respectively. The Berry curvature dipole (in addition to Berry curvature) was determined using WannierBerri package for a dense grid of 120 x 120 x 3 with an adaptive refinement mesh of 30 x 30 x 1. Frequency-dependent NLH current and intensity of emitted radiation of $\mathrm{WTe_2}$ monolayers and aligned gr-hBN were calculated using Mathematica software \cite{Wolfram}. Gaussian smearing width ($\mathit{\alpha}$) of the intensity peaks of emitted radiation in response to an incident light of frequency $\mathit{\omega_L\geqslant\omega_\mathrm{0}}$ was chosen based on the full width half maximum of the optical phonon linewidth ($\mathit{\Gamma_{ph}}$) observed in experiments \cite{Kim2016,Nemanich1981}.
	\subsection{Nonlinear Hall response from symmetry-permitted Berry curvature dipole}
		The applied electric field $\mathit{E_x(t)=\displaystyle{Re\left\{\mathcal{E_\mathit{x}}e^{i\omega_{\mathrm{AC}}t}\right\}}}$ shifts the Fermi surface in k-space (Fig. S1), giving rise to second order nonlinear Hall responses in a time-reversal invariant crystal as follows:
		\begin{eqnarray}
			j^{(\mathrm{0})}=-\varepsilon_{abc}\displaystyle{\frac{e^3\tau}{2(1+i\omega\tau)}\mathit{D_{bd}\mathcal{E_\mathit{x}^{(\omega_{\mathrm{AC}})}\mathcal{E_\mathit{x}^{*(\omega_{\mathrm{AC}})}}}}}, \\ j^{(\mathrm{2}\mathit{\omega})}=-\varepsilon_{abc}\displaystyle{\frac{e^3\tau}{\mathrm{2}(\mathrm{1}+i\omega\tau)}\mathit{D_{bd}\mathcal{E_\mathit{x}^{(\omega_{\mathrm{AC}})}}\mathcal{E_\mathit{x}^{(\omega_{\mathrm{AC}})}}}}
		\end{eqnarray}
		where  $\mathit{e}$ is the electronic charge and $\mathit{\tau}$ is the electron relaxation time. 
		\begin{subequations}
			\begin{align}
				D_{bd}&=\displaystyle{\int \Big[\frac{d^\mathrm{2}\bf{k}}{(\mathrm{2}\pi)^2}\Big]\mathit{d_{bd}\!~(\bf{k})}} \\
				&=\displaystyle{\sum_n \int \Big[\frac{d^\mathrm{2}\bf{k}}{(\mathrm{2}\pi)^2}\Big]\mathit{f\!~(\varepsilon_{\bf{k}})}\Big[\partial_{k_b} \mathit{\Omega_d\!~(\bf{k})}\Big]} \\
				&=\displaystyle{-\sum_n \int \Big[\frac{d^\mathrm{2}\bf{k}}{(\mathrm{2}\pi)^2}\Big]\frac{\partial \mathit{f\!~(\varepsilon_{\bf{k}})}}{\partial \varepsilon}v_b\!~(\bf{k}) \mathit{\Omega_d\!~(\bf{k})}}
			\end{align}
		\end{subequations}
		Berry curvature dipole $\mathit{D_{bd}}$, which is the first-order moment of $\mathit{\Omega_d\!~(\bf{k})}$, includes the summation of group velocity term from the band dispersion $\mathit{v_b\!~(\bf{k})=\displaystyle{\frac{\partial\varepsilon}{\partial \mathit{k_b}}}}$ over all the bands crossing the Fermi energy, as constrained by the Fermi-Dirac distribution, $\mathit{\displaystyle{\frac{\partial f\!~(\varepsilon_{\bf{k}})}{\partial \varepsilon}}}$. $\mathit{d_{bd}\!~(\bf{k})}$ is the $\mathit{D}$ density, which needs to be even under the symmetries of a 2D time-reversal invariant crystal for its $\mathit{D}$ to be nonzero (Table. \ref{S1}).
		
\begin{figure}[hbt!]
		\renewcommand{\thefigure}{S2}
		\includegraphics[width=17.8cm]{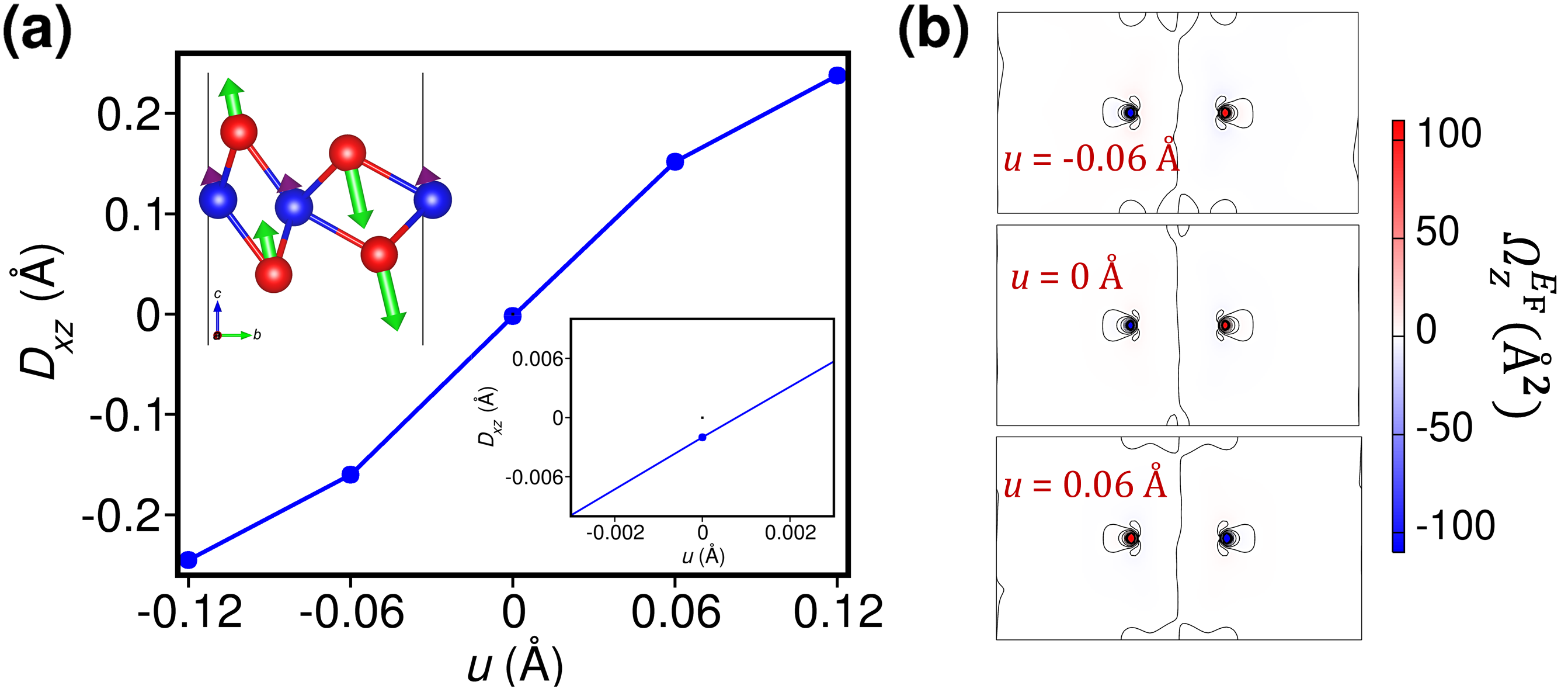}
		\caption{\label{TdWTe2} \textbf{GQuES-active phonon induced Berry curvature dipole of non-centrosymmetric $\mathit{T_d}$$\mathrm{-WTe_2}$ monolayer.} $\mathit{A\sp{\prime}}$ phonon of non-centrosymmetric $\mathit{T_d}$$\mathrm{-WTe_2}$ monolayer (inset) preserves $\mathit{\sigma_x}$ inherent to the monolayer and induces a change in $\mathit{D}$ that varies linearly with $\mathit{u}$ (a). Inset in (a) show inherent $\mathit{D}$ of $\mathit{T_d}$$\mathrm{-WTe_2}$ at $\mathit{u}$=0 (no distortion). Contour plots of $\mathit{\Omega_z\!~(\bf{k})}$ at $\mathit{E_\mathrm{F}}$ of $\mathit{T_d}$$\mathrm{-WTe_2}$ monolayer (b) show symmetry-allowed finite $\mathit{\Omega_z}$ at $\mathit{u}$=0, and how it changes with structural distortion ($\mathit{u<}$0 and $\mathit{u>}$0).}
	\end{figure}

		$\mathit{B_u}$ modes of $\mathit{T\sp{\prime}}$$\mathrm{-WTe_2}$ monolayer (Table. \ref{S2}) breaks the inherent inversion symmetry and leads to non-zero $\mathit{D}$, while the non-centrosymmetric $\mathit{T_d}$$\mathrm{-WTe_2}$ monolayer with a single mirror symmetry $\mathit{\sigma_x}$ exhibits an inherent $\mathit{D_{xz}}$, which is orthogonal to the $\mathit{\sigma_x}$ mirror line. The distortions induced by its amplitude of $\mathit{A\sp{\prime}}$ vibrational mode at $\mathit{\omega_\mathrm{0}}$=3.7 THz preserves $\mathit{\sigma_x}$ symmetry of $\mathit{T_d}$$\mathrm{-WTe_2}$ monolayer (Table. \ref{S3}) and induces changes in its $\mathit{D}$, as a consequence of change in the polarity of $\mathit{\Omega_z\!~(\bf{k})}$ (Fig. \ref{TdWTe2}). Due to the spin orbit coupling (SOC), electronic structure of $\mathit{T_d}$$\mathrm{-WTe_2}$ monolayer (at $\mathit{u}$=0, no distortion) exhibits a weak but non-zero $\mathit{D}$, as a consequence of the broken inversion symmetry. $\mathit{A\sp{\prime}}$ and $\mathit{A^{\prime\prime}}$ vibrational modes of $\mathit{T_d}$$\mathrm{-WTe_2}$ monolayer shows nonlinear Hall response at frequencies $\mathit{\omega_\mathrm{0} \pm \mathrm{2}\omega_{AC}}$, as a consequence of the presence or absence of a single mirror symmetry in the non-centrosymmetric $\mathit{T_d}$$\mathrm{-WTe_2}$ monolayer respectively. 
		
		Doubly degenerate $\mathit{E}$ vibrational modes of aligned gr-hBN results in nonzero $\mathit{D}$, as a consequence of $\mathit{C_{3z}}$ symmetry breaking (Table. \ref{S4}). One of the $\mathit{E}$ modes break $\mathit{C_{3z}}$ symmetry leading to nonzero $\mathit{D_{xz}}$ orthogonal to $\mathit{\sigma_x}$ mirror symmetry, while the other breaks both $\mathit{C_{3z}}$ and $\mathit{\sigma_x}$ symmetries, resulting in nonzero $\mathit{D_{xz}}$ and $\mathit{D_{yz}}$ (Fig. S5). 
		
	\begin{table}[hbt!]
		\renewcommand{\thetable}{S1}
		\caption{\label{S1}Symmetries of a 2D crystal that permits its $\mathit{D}$ to be nonzero and constrain it to be zero}
		\begin{ruledtabular}
			\begin{tabular}{ccc}
				Crystalline symmetry & $\mathit{d_{bz}\!~(\bf{k})}$ & $\mathit{D_{bz}}$ \\
				\hline
				Time-reversal & $\mathit{d_{xz}\!~(\bf{k})}$,$\mathit{d_{yz}\!~(\bf{k})}$ & $\mathit{D_{xz}}$,$\mathit{D_{yz}}$ \\
				Mirror ($\mathit{\sigma_x}$) & $\mathit{d_{xz}\!~(\bf{k})}$ & $\mathit{D_{xz}}$ \\
				Mirror ($\mathit{\sigma_y}$) & $\mathit{d_{yz}\!~(\bf{k})}$ & $\mathit{D_{yz}}$ \\
				Rotation ($\mathit{C_{2z}}$) & 0 & 0 \\
				Inversion ($\mathit{\hat{I}}$) & 0 & 0 
			\end{tabular}
		\end{ruledtabular}
	\end{table}
	
	\begin{table}[hbt!]
		\renewcommand{\thetable}{S2}
		\caption{\label{S2}GQuES-active phonons of $\mathit{T\sp{\prime}}$$\mathrm{-WTe_2}$ monolayer that permits its $\mathit{D}$ to be nonzero}
		\begin{ruledtabular}
			\begin{tabular}{ccccc}
				$\mathit{T\sp\prime}$-$\mathrm{WTe_2}$ monolayer & $\mathit{E}$ & $\mathit{C_{2x}}$ & $\mathit{\hat{I}}$ & $\mathit{\sigma_x}$ \\
				\hline
				$\mathit{B_u}$ & 1 & -1 & -1 & 1 
			\end{tabular}
		\end{ruledtabular}
	\end{table}
	
	\begin{table}[hbt!]
		\renewcommand{\thetable}{S3}
		\caption{\label{S3}GQuES-active phonons of $\mathit{T_d}$$\mathrm{WTe_2}$ monolayer that permits its $\mathit{D}$ to be nonzero}
		\begin{ruledtabular}
			\begin{tabular}{ccc}
				$\mathit{T_d}$-$\mathrm{WTe_2}$ monolayer & $\mathit{E}$ & $\mathit{\sigma_x}$ \\
				\hline
				$\mathit{A\sp{\prime}}$ & 1 & 1 \\
				$\mathit{A^{\prime\prime}}$ & 1 & -1 
			\end{tabular}
		\end{ruledtabular}
	\end{table}
	
	\begin{table}[hbt!]
		\renewcommand{\thetable}{S4}
		\caption{\label{S4}GQuES-active phonons of aligned gr-hBN that permits its $\mathit{D}$ to be nonzero}
		\begin{ruledtabular}
			\begin{tabular}{cccc}
				Aligned gr-hBN & $\mathit{E}$ & 2$\mathit{C_{3z}}$ & 3$\mathit{\sigma_x}$ \\
				\hline
				$\mathit{E}$ & 2 & -1 & 0 
			\end{tabular}
		\end{ruledtabular}
	\end{table}
	\begin{figure}[hbt!]
		\renewcommand{\thefigure}{S1}
		\includegraphics[width=17.8cm,keepaspectratio]{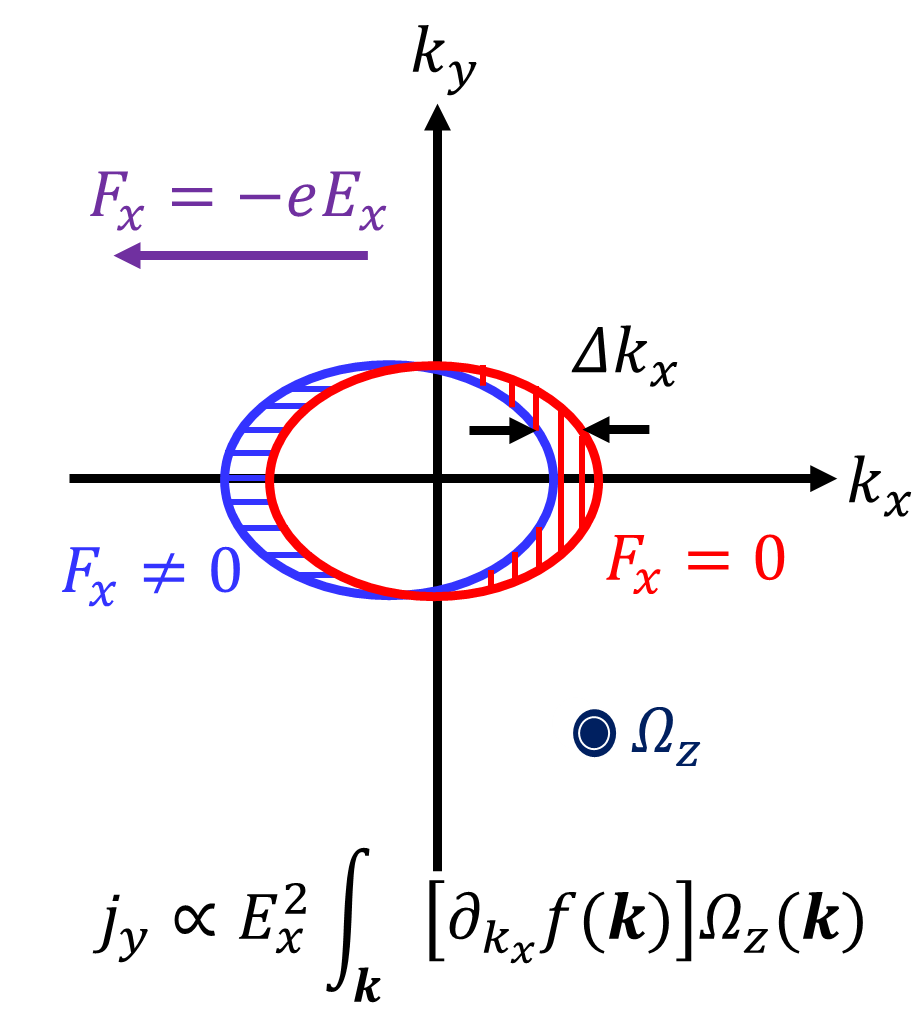}
		\caption{\label{NLHresp} \textbf{Origin of nonlinear Hall response.} Shift in the Fermi surface along negative $\mathit{k_x}$ direction, under the influence of $\mathit{F_x}$ generated by an in-plane electric field $\mathit{E_x}$ along positive $\mathit{\hat{x}}$ direction leads to second-order nonlinear Hall current $\mathit{j_y}$, as a consequence of non-equilibrium distribution of electrons given by $\mathit{\partial_{k_x}\/ f\!~(\bf{k})}$.}
	\end{figure}
	
	\newpage
	
	\begin{figure}[hbt!]
		\renewcommand{\thefigure}{S3}
		\includegraphics[width=17.8cm]{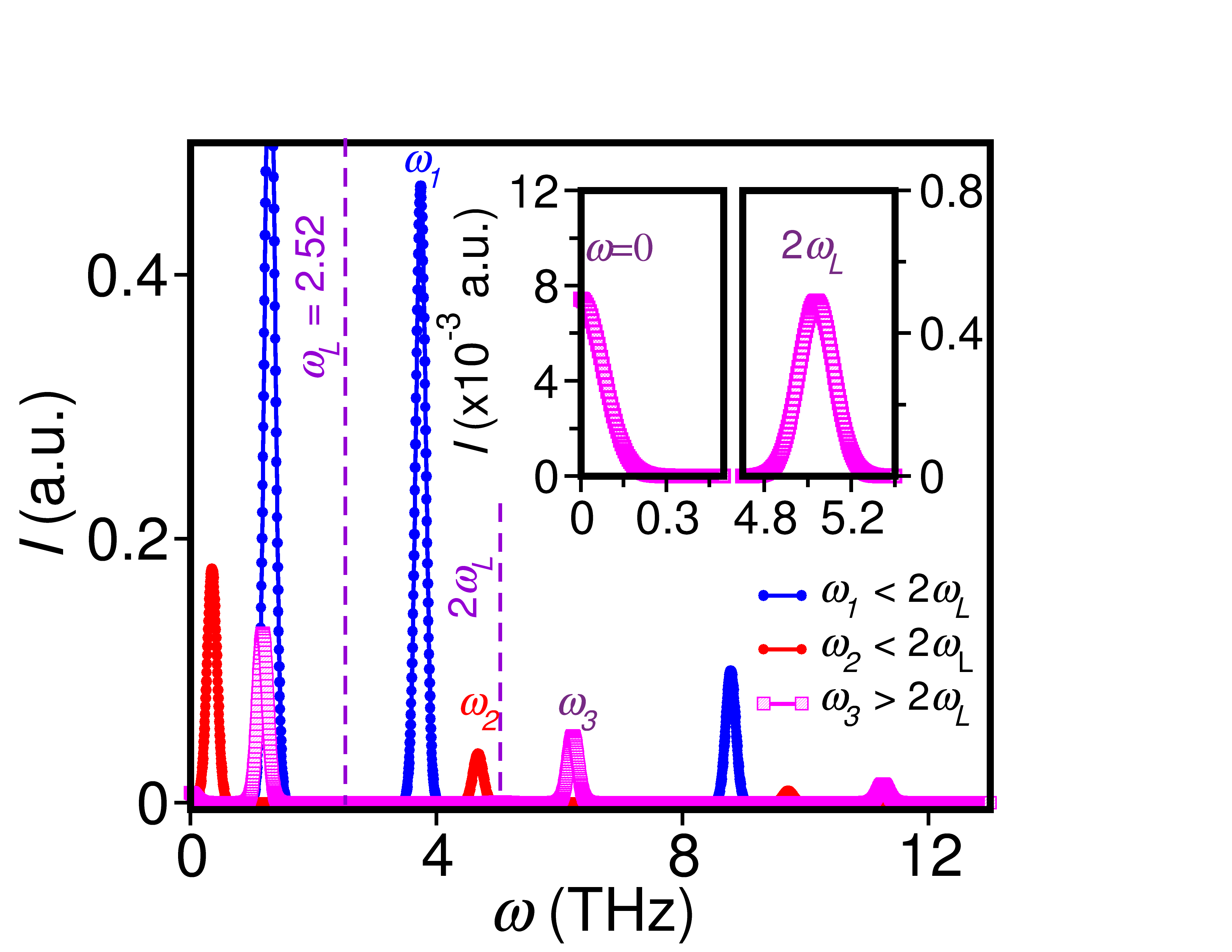}
		\caption{\label{freqTd} \textbf{Predictions for experimental observations of GQuES spectrum of $\mathit{T_d}$$\mathrm{-WTe_2}$ monolayer.} In response to incident light at frequency $\mathit{\omega_L}$=2.52 THz (laser wavelength of 118.9 $\mathit{\mu}$m), some of all ($\mathit{A\sp{\prime}}$ and $\mathit{A^{\prime\prime}}$) GQuES-active phonons ($\mathit{\omega_\mathrm{1}}$=3.75 THz, $\mathit{\omega_\mathrm{2}}$=4.68 THz, $\mathit{\omega_\mathrm{3}}$=6.22 THz) of $\mathit{T_d}$$\mathrm{-WTe_2}$ monolayer are evident in peaks at $\mathit{\omega=\mathrm{2}\omega_L \pm \omega_\mathrm{0}}$ (blue and red solid circles) and at $\mathit{\omega=\omega_\mathrm{0} \pm \mathrm{2}\omega_L}$ (magenta squares). Additionally, peaks at $\mathit{\omega}$=0 and at 2$\mathit{\omega_L}$ (5.04 THz) are seen in the insets.}
	\end{figure}
	
	\begin{figure}[hbt!]
		\renewcommand{\thefigure}{S4}
		\includegraphics[width=17.8cm]{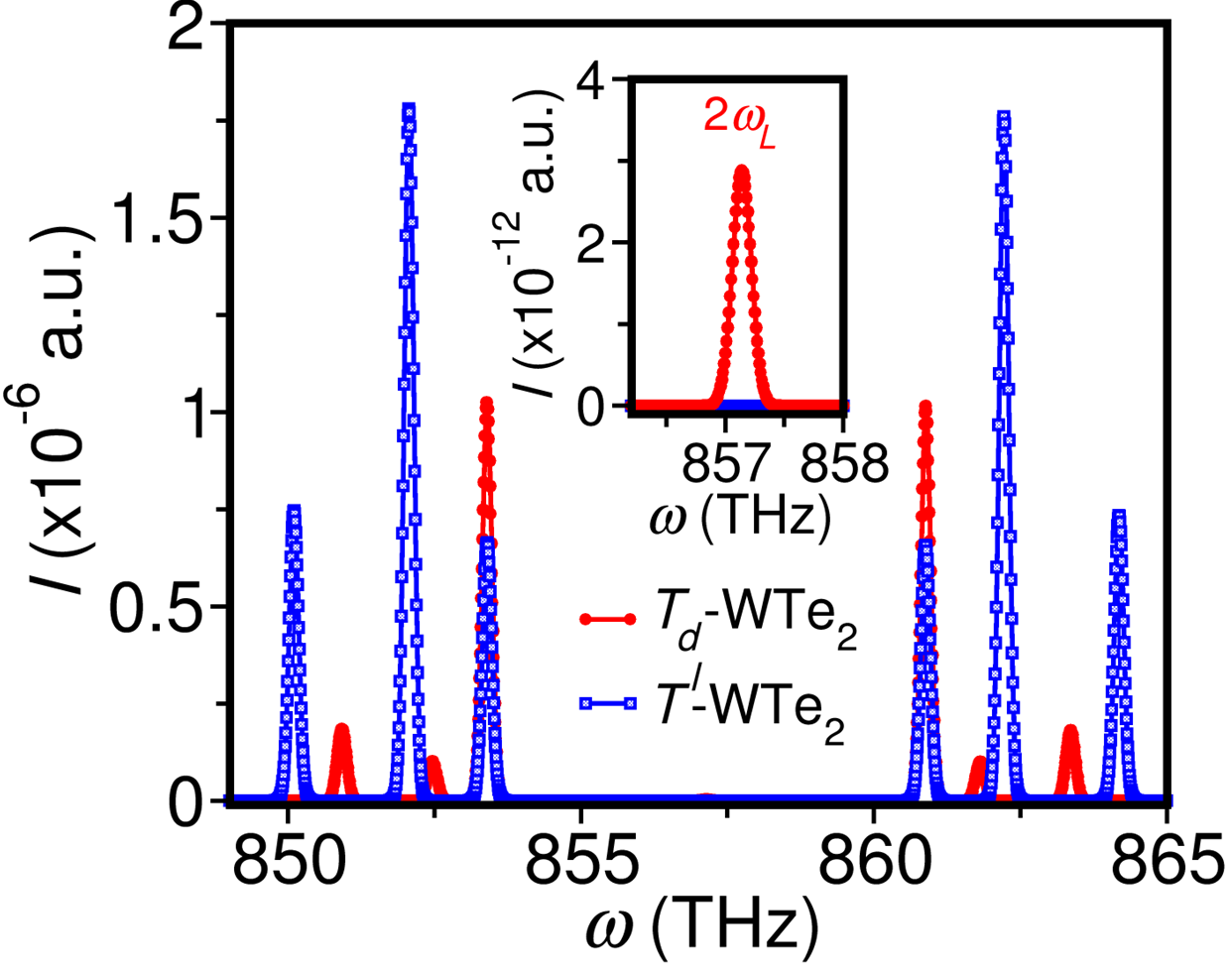}
		\caption{\label{optfreq} \textbf{Predictions of GQuES spectrum of $\mathrm{WTe_2}$ monolayer at optical frequencies.} In response to incident light at frequency $\mathit{\omega_L}$=428.6 THz (laser wavelength of 700 nm), GQuES-active phonons of $\mathit{T_d}$$\mathrm{-WTe_2}$ and $\mathit{T\sp{\prime}}$$\mathrm{-WTe_2}$ monolayer are evident in peaks centred at 2$\mathit{\omega_L}$. Additionally, intrinsic $\mathit{D}$ of $\mathit{T_d}$$\mathrm{-WTe_2}$ monolayer results in a peak at 2$\mathit{\omega_L}$ (857.2 THz) and at $\mathit{\omega}$=0 (not shown here).}
	\end{figure}
	
	\begin{figure}[hbt!]
		\renewcommand{\thefigure}{S5}
		\includegraphics[width=17.8cm]{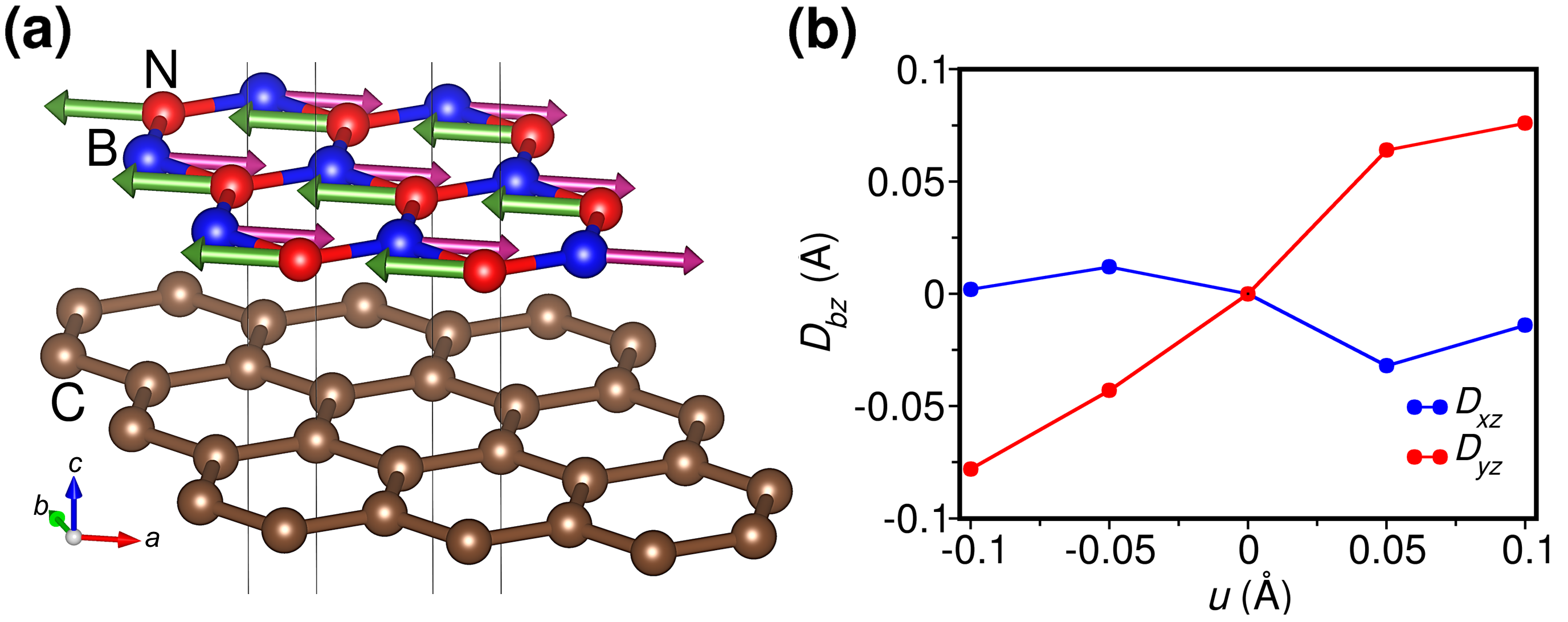}
		\caption{\label{grhBN} \textbf{GQuES-active phonon of aligned gr-hBN.} Normal mode displacements of one of the doubly degenerate $\mathit{E}$ phonons at $\mathit{\omega_\mathrm{0}}$=41.9 THz of aligned gr-hBN involve distortions primarily in hBN along cartesian $\mathit{\hat{x}}$ direction (a) breaking $\mathit{C_{3z}+\sigma_x}$ symmetries, thereby allowing $\mathit{D_{bz} (b=x,y)}$ Dbz to be nonzero (b).}
	\end{figure}
	\FloatBarrier